\documentclass[epj]{svjour}

\usepackage{graphicx}

\def \be {\begin{equation}}
\def \ee {\end{equation}}
\def \bea {\begin{eqnarray}}
\def \eea {\end{eqnarray}}

\begin{document}

\title{ Effects of acceleration on the
collision of particles in the rotating black hole spacetime}

\author{Weiping Yao$^{1}$\and Songbai Chen$^{1,}$\thanks{\emph{E-mail:} csb3752@163.com}
\and Changqing Liu$^{1,2}$ \and Jiliang Jing
$^{1,}$\thanks{\emph{E-mail:} jljing@hunnu.edu.cn}} \institute{1
Institute of Physics and Department of Physics, Hunan Normal
University, Changsha, Hunan 410081, P. R. China.\\Key Laboratory of
Low Dimensional Quantum Structures and Quantum Control (Hunan Normal
University), Ministry
of Education, P. R. China.\\
2 Department of Physics and Information Engineering, Hunan Institute
of Humanities Science and Technology, Loudi, Hunan 417000, P. R.
China.}

\date{Received: date / Revised version: date}

\abstract{ We study the collision of two geodesic particles in the
accelerating and rotating black hole spacetime and probe the effects
of the acceleration of black hole on the center-of-mass energy of
the colliding particles and on the high-velocity collision belts. We
find that the dependence of the center-of-mass energy on the
acceleration in the near event-horizon collision  is different from
that in the near acceleration-horizon case. Moreover, the presence
of the acceleration changes the shape and position of the
high-velocity collision belts. Our results show that the
acceleration of black holes brings richer physics for the collision
of particles.}

 \maketitle

\section{Introduction}
\label{intro}

Recently, Banados, Silk and West (BSW) \cite{BSW2009} investigated
the collision of two geodesic particles near the horizon of an
extremal rotating black hole \cite{Kerr1963} and found that the
rotating black hole could be regarded as a Planck-energy-scale
collider since the center-of-mass (CM) energy for a pair particles
can be unlimited if the angular momentum of either particle $L_i$ is
fine-tuned to the critical angular momentum $L_C$. This means that
we might obtain some visible signals about ultra-high energy physics
through such a BSW mechanism about the collision of two particles
near a rotating black hole. Therefore, a lot of efforts have been
focused on the BSW mechanism in the recent years
\cite{Berti,Ted1,Lake2010,Grib1,Grib2,Grib3,Grib4,MB1,WLGF1,WLGF2,PJ,OB1,OB2,OB3,Mki,MB2,Andrew,Liu1,Ja,MP1,MP2,MP3,WS1,WS2}.
It was pointed out \cite{Berti,Ted1} that the arbitrarily high CM
energy $E_{cm}$ for a Kerr black hole could not be achievable in
nature due to the astrophysical limitation, such as the maximal spin
and gravitational radiation. Subsequently, Lake \cite{Lake2010} also
found the CM energy is limited as the collision occurring at the
inner horizon of the non-extremal Kerr black hole. Grib and Pavlov
\cite{Grib1,Grib2,Grib3,Grib4} argued that the CM energy $E_{cm}$
for two particles collision can be unlimited even in the non-maximal
rotation if one considers the multiple scattering, and they also
evaluated extraction of energy after the collision. Wei \textit{et
al.} \cite{WLGF1,WLGF2} studied the effects of the charge on the CM
energy for the colliding particles in the stringy black hole and
Kerr-Newman black holes. Banados \textit{et al.} \cite{MB2} and
Williams \cite{Andrew} calculated the escaping flux of massless
particles for maximally rotating black holes, which shows that the
received spectrum should contain some typical signatures of highly
energetic products. Zaslavskii \cite{OB1} discussed the universal
property of the collision of particles for a rotating black hole and
tried to give a general explanation of this BSW mechanism. Moreover,
Zaslavskii \textit{et al} \cite{OB2} also found the similar BSW
process for the colliding charged particles in the non-rotating
charged black hole spacetime. The similar investigations have also
been done in \cite{WS1,WS2}.

However, all of the works above have been focused on the colliding
particles moving in the equatorial plane. Recently, Harada
\cite{TM1} considered two colliding general geodesic massive and
massless particles at any spacetime point around a Kerr black hole
and obtained an explicit expression for the CM energy of the
particles. He found that the collision with an unboundedly high CM
energy can occur only in the belt centered at the equator of an
extremal rotating black hole. In Ref.\cite{Liu2}, we extended
Harada's work to the Kerr-Newman black hole and obtained that the
charge $q$ decreases the value of the latitude of the high-velocity
collision belts in which arbitrarily high CM energy can occurs.
These results can help us to understand more about the BSW process
in the rotating black holes.

In this paper, we will investigate the collision of two particles in
the accelerating and rotating black holes spacetime
\cite{Hong.k2005,J.B.Griffiths2005,J.B.Griffiths2006}, which
describes a pair of causally separated black holes accelerating away
from each other under the action of `` strings " represented by
conical singularities located along appropriate sections of the axis
of symmetry. The presence of the acceleration results in that the
spacetime has different geometric structure from that of the usual
Kerr case. The accelerating and rotating black holes possess two
rotation horizons and two acceleration horizons, which implies that
both Hawking and Unruh radiation could be present in this
background. The properties of the accelerating and rotating black
holes have been studied extensively in recent years
\cite{Hong.k2005,J.B.Griffiths2005,J.B.Griffiths2006,Hawking,JM2,US2,JH1}.
Saifullah \textit{et al} \cite{JM2,US2} discussed the surface
gravity, Hawking temperature and the area laws for accelerating and
rotating black holes and studied the Hawking radiation of scalar
particles in these black holes spacetime. Hawking and Ross
\cite{Hawking} used such a kind of metric to probe the possible
creation of a black hole pair by the breaking of a cosmic string.
Thus, the study of the accelerating and rotating black holes could
provides physical insight into the high energy physics. The main
purpose of this paper is to see how the acceleration affects the CM
energy and the high-velocity collision belts for the colliding
particles in the accelerating and rotating black holes spacetime.

The paper is organized as follows. In Sec. II, we review briefly the
accelerating and rotating black hole spacetime and discuss the
geodesic motion for a particle in this background. In Sec. III, we
investigate the collision of two massive geodesic particles in the
accelerating and rotating black hole spacetime and probe the effects
of the acceleration of black hole on the CM energy of the colliding
particles. In Sec.IV, we study the effects of the acceleration on
the high-velocity collision belts for the colliding particles in the
accelerating and rotating black hole spacetime. We end the paper
with a summary.

\section{Geodesic motion of a particle in the
accelerating and rotating black hole spacetime}

Let us now review briefly the accelerating and rotating black hole
spacetime. The Pleba\'{n}ski and Demia\'{n}ski metric \cite{JM1}
covers a large family of electro-vacuum type-$D$ spacetimes, which
includes both the Kerr-Newman-like solutions and the $C$-metric. The
accelerating and rotating black hole is a special case of this
family, which describes the gravitational field by a pair of
uniformly accelerating Kerr-type black holes. In the Boyer-Lindquist
coordinates, the metric of this accelerating and rotating black
holes spacetime can be expressed as
\cite{Hong.k2005,J.B.Griffiths2005,J.B.Griffiths2006}
\begin{eqnarray}
ds^{2} &=&-\left( \frac{\Delta-a^{2}P\sin ^{2}\theta }{\rho
^{2}\Omega ^{2}} \right) dt^{2}+\frac{\rho ^{2}dr^{2}}{\Delta\Omega
^{2}}+
\frac{\rho ^{2}d\theta ^{2} }{P\Omega ^{2}} \nonumber \\
&&+\left( \frac{\sin ^{2}\theta \left[ P\left( r^{2}+a^{2}\right)
^{2}-a^{2}\Delta\sin ^{2}\theta \right] }{\rho ^{2}\Omega
^{2}}\right) d\phi ^{2}
\nonumber \\
&&-\left( \frac{2a\sin ^{2}\theta \left[ P\left( r^{2}+a^{2}\right)
-\Delta\right] }{\rho ^{2}\Omega ^{2}}\right) dtd\phi, \label{1}
\end{eqnarray}
with
\begin{eqnarray}
\Omega &=&1-\alpha r\cos \theta , \label{1.1} \\
\rho ^{2} &=&r^{2}+a^{2}\cos ^{2}\theta ,  \label{1.2} \\
P &=&1-2\alpha M\cos \theta + \alpha ^{2}
a^{2}\cos ^{2}\theta ,  \label{1.3} \\
\Delta &=&(r^{2}-2Mr+a^{2})(1-\alpha ^{2}r^{2}) . \label{1.4}
\end{eqnarray}
Here $M$ is the mass of the black hole, $\alpha$ is the acceleration
of the black hole and $a$ is angular momentum per unit mass. The
horizons of this black hole can be calculated by putting $g^{rr}=0$
\cite{Hong.k2005,J.B.Griffiths2005,J.B.Griffiths2006}, i.e.,
\begin{eqnarray}
\frac{\Delta \Omega^2}{\rho^2}=0,
\end{eqnarray}
which gives their location at
\begin{eqnarray}
 r_{H }&=&M+
 \sqrt{M^{2}-a^{2}},\;\;\;\;\;r_{C}=M-\sqrt{M^{2}-a^{2}},\nonumber\\
 r_A&=&\frac{1}{\alpha },\;\;\;\;r_{\alpha}=\frac{1}{\alpha \cos\theta}. \label{3}
\end{eqnarray}
Obviously, the position of the event horizon $r=r_H$ and Cauchy
horizon $r=r_{C}$ are same to those of the Kerr black hole. However,
in this case, there exist the other horizons at $r_A=\frac{1}{\alpha
}$ and $r_{\alpha}=\frac{1}{\alpha \cos\theta}$, which is already
familiar in the context of the C-metric as the acceleration
horizons. The physical region of the black hole is $r_H<r<r_A$ in
which $\Delta>0$.

The non-vanishing components of the inverse metric for the black
 hole (\ref{1}) can be expressed as
\begin{eqnarray}
g^{tt}&=&-\frac{\Omega^{2}[P(r^{2}+a^{2})^{2}-a^{2}\Delta\sin^{2}\theta]}{\rho^{2}P\Delta},
\;\;\;\;g^{rr}=\frac{\Delta\Omega^{2}}{\rho^{2}},
\nonumber \\
g^{t\phi}&=&g^{\phi
t}=-\frac{\Omega^{2}a[P(r^{2}+a^{2})-\Delta]}{\rho^{2}P\Delta},\;\;\;\;\;\;g^{\theta\theta}=\frac{\Omega^{2}P}{\rho^{2}},
\nonumber \\
g^{\phi\phi}&=&\frac{\Omega^{2}(\Delta-a^{2}P\sin^{2}{\theta})}{\rho^{2}P\Delta\sin^{2}\theta}.
\label{1.5}
\end{eqnarray}
The Hamiltonian for the geodesic motion in the general curve
spacetime can be expressed as \cite{Cart1}
\begin{equation}
{\cal H}[x^{\alpha},p_{\beta}]=\frac{1}{2}\sum_{\mu,\nu}
g^{\mu\nu}p_{\mu}p_{\nu},\label{JHo1}
\end{equation}
where $p_{\alpha}$ is the conjugate momentum to $x^{\alpha}$. Since
the action $S=S(\lambda, x^{\alpha})$ is a function of the parameter
$\lambda$ and coordinates $x^{\alpha}$,  the conjugate momentum
$p_{\alpha}$ can be described by $p_{\alpha}=\partial S/\partial
x^{\alpha}$, and then the corresponding Hamilton-Jacobi equation is
given by \cite{Cart1}
\begin{equation}
-\frac{\partial S}{\partial \lambda}={\cal
H}\left[x^{\alpha},\frac{\partial S}{\partial x^{\beta}}\right]
=\frac{1}{2}\sum_{\mu,\nu}g^{\mu\nu}\frac{\partial S}{\partial
x^{\mu}}\frac{\partial S}{\partial x^{\nu}}.\label{eq:J H0}
\end{equation}
In the accelerating and rotating black hole spacetime,
 the Hamilton-Jacobi equation for the  geodesic particle can be written as
\begin{eqnarray}
&&-\frac{\partial S}{\partial
\lambda}=\frac{\Omega^{2}}{2\rho^{2}}\left\{-\frac{1}{\Delta}
\left[(r^{2}+a^{2})\frac{\partial S}{\partial t}+a \frac{\partial
S}{\partial \phi}\right]^{2}
+\Delta\left(\frac{\partial S}{\partial r}\right)^{2}\right. \nonumber \\
&&+\left.\frac{1}{P\sin^{2}\theta}\left[\frac{\partial S}{\partial
\phi} +a\sin^{2}\theta \frac{\partial S}{\partial t}\right]^{2}
+P\left(\frac{\partial S}{\partial \theta}\right)^{2}\right\}.
\label{eq:HJ_explicit}
\end{eqnarray}
Since $\lambda$, $t$ and $\phi$ are cyclic coordinates,  the action
$S$ has the form
\begin{equation}
S=\frac{1}{2}m ^{2}\lambda-Et+L\phi+S_{r}(r)+S_{\theta}(\theta),
\label{eq:J H}
\end{equation}
where $m$, $E$ and $L$ are constants which correspond to the rest
mass, conserved energy and angular momentum of the particle.
Substituting Eq. (\ref{eq:J H}) into Eq. (\ref{eq:HJ_explicit}), we
find that the Hamilton-Jacobi equation can be simplified as
\begin{eqnarray}
&&-m^{2}\rho^{2}=\Omega^{2}\left\{-\frac{[(r^{2}+a^{2})E-aL]^{2}}{\Delta}
+\Delta\left(\frac{dS_{r}}{dr}\right)^{2}\right.\nonumber\\
&&
\left.+\frac{1}{P\sin^{2}\theta}\bigg(L-aE\sin^{2}\theta\bigg)^{2}
+P\left(\frac{dS_{\theta}}{d\theta}\right)^{2}\right\}.
\label{eq:SrSth}
\end{eqnarray}
It is clear that for the massless particle the equation
(\ref{eq:SrSth}) can be separable. However, the collision between
massless particles does not occur really since in the regular point
two ingoing null geodesics cannot intersect. For the massive
particle, we find that the equation (\ref{eq:SrSth}) is not
separable generally  because the factor $\Omega^{2}$ in the
right-hand side of Eq.(\ref{eq:SrSth}) is a function of the product
of $r$ and $\cos\theta$. Here, we assume that the collision of
massive particles occurs in the region $r\ll
|\frac{1}{\alpha\cos{\theta}}|$ so that $\Omega^{2}\sim 1$, which
ensures the equation (\ref{eq:SrSth}) can be separable in this
limit. Thus, in this paper, we will focus only on the collision of
massive particles in the region $r\ll
|\frac{1}{\alpha\cos{\theta}}|$, and then probe the effects of the
acceleration on the CM energy and on the allowed range of the value
of $\theta$ for the particles collision in the accelerating and
rotating black hole spacetime.

\section{Collision of two massive geodesic particles
in the accelerating and rotating black hole spacetime}

In the region $r\ll |\frac{1}{\alpha\cos{\theta}}|$, we have
$\Omega^2\sim 1$, the Hamilton-Jacobi equation for a massive
particle can be approximated as
\begin{eqnarray}
&&\Delta\left(\frac{dS_{r}}{dr}\right)^{2}-\frac{[(r^{2}+a^{2})E-aL]^{2}}{\Delta}+m^2r^2+K=0,
\nonumber\\\label{eqKr}
\end{eqnarray}
and
\begin{eqnarray}
P\left(\frac{dS_{\theta}}{d\theta}\right)^{2}
+\frac{1}{P\sin^{2}\theta}[L-aE\sin^{2}\theta]^{2}&+&m^2a^2 \cos^2{\theta}\nonumber\\
&-&K=0.\label{eqKth}
\end{eqnarray}
$K$ is a Carter-like constant, which is related to black hole's
parameters ($M$, $a$, $\alpha$) and the particle's parameters ($E$,
$L$, $m$). Integrating Eqs.(\ref{eqKr}) and (\ref{eqKth}), we can
obtain
\begin{eqnarray}
S_{r}&=&\sigma_{r}\int^{r}\frac{\sqrt{R(r)}}{\Delta}dr,\label{sr}\\
S_{\theta}&=&\sigma_{\theta}\int^{\theta}d\theta\frac{\sqrt{\Theta(\theta)}}{P},
\end{eqnarray}
with
\begin{eqnarray}
&&R(r)=\mathcal{P}(r)^{2}-\Delta(r)(m^2r^2+K),
\label{eqR}\\
&&\mathcal{P}(r)=(r^{2}+a^{2})E-aL,
\label{eqTheta}\\
&&\Theta(\theta)=(K-m^2a^2\cos^{2}\theta)P-
\bigg(\frac{L-aE\sin^{2}\theta}{\sin\theta}\bigg)^2,\nonumber\\
\label{eqThetaRP}
\end{eqnarray}
where $\sigma_{r}$ and $\sigma_{\theta}$ are two independent sign
functions. Using Eqs.(\ref{JHo1}), (\ref{eq:J H}) and
(\ref{sr})-(\ref{eqTheta}), we can obtain the geodesic equations of
the particle in the accelerating and rotating black hole spacetime
\begin{eqnarray}
\rho^{2}\frac{dt}{d\lambda}&=&\bigg[-\frac{a(aE\sin^{2}\theta-L)}{P}+
\frac{(r^{2}+a^{2})\left((r^{2}+a^{2})E-aL\right)}{\Delta}\bigg],\nonumber\\\label{eqdtdlambda}\\
\rho^{2}\frac{dr}{d\lambda}&=&\sigma_{r} \sqrt{R}, \label{eqdrdlambda}\\
\rho^{2}\frac{d\theta}{d\lambda}&=&\sigma_{\theta} \frac{\sqrt{\Theta}}{P}, \label{eqdthetadlambda}\\
\rho^{2}\frac{d\phi}{d\lambda}&=&\bigg[-\frac{1}{P}\bigg(a
E-\frac{L}{\sin^{2}\theta}\bigg)+\frac{a\left((r^{2}+a^{2})E-aL\right)}{\Delta}\bigg].\label{eqdphiambda}
\end{eqnarray}
Then, the radial equation for the massive particle  moving along
geodesics can be expressed as
\begin{equation}
\frac{1}{2}\left(\frac{dr}{d\lambda}\right)^{2}+V_{\rm eff}(r)=0,
\label{eq:mechanical_energy1}
\end{equation}
with the effective potential
\begin{equation}
V_{\rm eff}(r)\equiv -\frac{ R(r)}{2\rho^{4}}.
\label{eq:effective_potential1}
\end{equation}
The circular orbit of the particle satisfies the conditions
\begin{eqnarray}
V_{\rm eff}(r)=0,\;\;\;\;\;\;\frac{d~V_{\rm eff}(r)}{dr}=0
\label{eq:circular orbit1}.
\end{eqnarray}
Since the geodesics is timelike, we have $dt/d\lambda\geq 0$, which
means
\begin{eqnarray}
-\frac{a(aE\sin^{2}\theta-L)}{P}
+\frac{(r^{2}+a^{2})\left((r^{2}+a^{2})E-aL\right)}{\Delta}\geq0.\nonumber\\
\end{eqnarray}
From Eq.(\ref{1.3}), one can find that the quantity $P$ can be
rewritten as $P=\alpha^2(r_A-r_H\cos\theta)(r_A-r_C\cos\theta)$,
which ensures $P>0$ in the physical region of black hole. Thus, the
causal conditions for the particle are reduced to
\begin{eqnarray}
E\geq\frac{aL}{r_{H}^{2}+a^{2}}=\Omega_{H}L,\label{yg}
\end{eqnarray}
near the event horizon and
\begin{eqnarray}
E\geq\frac{aL}{r_{A}^{2}+a^{2}}=\Omega_{A}L,
\end{eqnarray}
near the acceleration horizon, respectively. When $\alpha\rightarrow
0$, one can obtain that the acceleration horizon $r_A\rightarrow
\infty$ and the critical angular momentum $L_{cA} \rightarrow
\infty$. This means that the particle with the arbitrarily angular
momentum $L$ at spatial infinity satisfies the causal condition in
the usual non-accelerating Kerr black hole. Moreover, we also find
that the critical value $L_{cH}$ is independent of the acceleration
$\alpha$ of the black hole.

Assuming that two massive geodesic particles 1 and 2 are at the same
spacetime point with the four momenta $p^a_{(i)}$,  we can obtain
that the CM energy $E_{cm}$ for the particles in the accelerating
and rotating black hole spacetime has the form \cite{TM1}
\begin{eqnarray}
E_{cm}^{2}&=&-p^{a}_{\rm tot}p_{{\rm tot}a}=m_{1}^{2}+m_{2}^{2}
-2g_{ab}p^{a}_{1}p_{2}^{b}\nonumber\\
&=&m_{1}^{2}+m_{2}^{2}+\frac{2}{\rho^{2}}\bigg[\frac{\mathcal{P}(r)_{1}\mathcal{P}(r)_{2}
-\sigma_{1r}\sqrt{R_{1}}\sigma_{2r}\sqrt{R_{2}}}{\Delta}\nonumber\\
&-&\frac{(L_{1} -a\sin^{2}\theta E_{1})(L_{2}-a\sin^{2}\theta
E_{2})}{P\sin^{2}\theta}\nonumber\\
&-&\frac{\sigma_{1\theta}
\sqrt{\Theta_{1}}\sigma_{2\theta}\sqrt{\Theta_{2}}}{P}\bigg],
 \label{mcm1}
\end{eqnarray}
where $m_i$ is  the rest mass of the particle $i$ ($i=1$, $2$). As
in Refs.\cite{TM1,Liu2}, we here consider only the case that two
colliding particles  have the same signs exactly. For the collision
of the particles near the event horizon, i.e., $r\sim r_{H}$, we
have the expression of CM energy of the particles
\begin{eqnarray}
E_{
cm_H}^{2}&=&m_{1}^{2}+m_{2}^{2}+\frac{1}{r^2_H+a^2\cos^2\theta}\bigg[(m^2_1r^2_H+K_1)
\nonumber\\&\times&\frac{E_{2}- \Omega_{H}L_{2}}{E_{1}-
\Omega_{H}L_{1}}+(m^2_2r^2_H+K_2)
\frac{E_{1}-\Omega_{H}L_{1}}{E_{2}-\Omega_{H}L_{2}}\nonumber\\&-&\frac{2(L_{1}-a\sin^{2}\theta
E_{1})(L_{2}-a\sin^{2}\theta E_{2})}{(1-2\alpha M\cos \theta
+\alpha^{2}a^{2} \cos
^{2}\theta)\sin^{2}\theta}\nonumber\\&-&\frac{2\sigma_{1\theta}\sqrt{\Theta_{1}}\sigma_{2\theta}\sqrt{\Theta_{2}}}{1-2\alpha
M\cos \theta +\alpha^{2}a^{2} \cos ^{2}\theta}\bigg].
 \label{eq:Omega_formula}
\end{eqnarray}
It shows that the unboundedly high CM energy can be obtained when
one of the particles has the critical angular momentum
$L_c=\frac{E}{\Omega_H}$, which is similar to that in the usual
non-accelerating Kerr black hole case. The derivatives of
$E^2_{cm_H}$ with respect to $\alpha$ can be expressed as
\begin{eqnarray}
\frac{dE^2_{
cm_H}}{d\alpha}&=&\frac{\cos\theta(a^2\alpha\cos\theta-M)}{P^2(r^2_H+a^2\cos^2\theta)}
\bigg\{\frac{2(L_{1}-a\sin^{2}\theta
E_{1})}{\sin^{2}\theta}\nonumber\\&\times&(L_{2}-a\sin^{2}\theta
E_{2})+\sigma_{1\theta}
\sqrt{\Theta_{1}}\sigma_{2\theta}\sqrt{\Theta_{2}}\bigg[2-P\nonumber\\&\times&
\bigg(\frac{K_1-m^2_1a^2\cos^2\theta}{\Theta_1}
+\frac{K_2-m^2_2a^2\cos^2\theta}{\Theta_2}\bigg)\bigg]
\bigg\}\nonumber\\&+&\frac{1}{r^2_H+a^2\cos^2\theta}\bigg[\bigg
(\frac{E_{1}-\Omega_{H}L_{1}}{E_{2}-\Omega_{H}L_{2}}-\sigma_{1\theta}\sigma_{2\theta}
\sqrt{\frac{\Theta_1}{\Theta_2}}\bigg)\nonumber\\&\times&\frac{dK_1}{d\alpha}+\bigg(
\frac{E_{2}-\Omega_{H}L_{2}}{E_{1}-\Omega_{H}L_{1}}-\sigma_{1\theta}\sigma_{2\theta}
\sqrt{\frac{\Theta_2}{\Theta_1}}\bigg)\frac{dK_2}{d\alpha}\bigg].\nonumber\\\label{ds1}
\end{eqnarray}
This formula tells us that the rate of change
$\frac{dE^2_{cm_H}}{d\alpha}$ depends not only on the parameters of
the black hole and of the colliding particles, but also on the polar
angle $\theta$ of the orbit of particles.  For particles located in
the equatorial plane, we can obtain that
$\frac{dE^2_{cm_H}}{d\alpha}=0$ and the CM energy is independent of
the acceleration of the black hole, which can be explained by a fact
that in the equatorial plane the Carter's constants $K_1$, $K_2$ do
not depend on the acceleration $\alpha$ and all of the terms
containing $\alpha$ in the metric functions $\Omega$ and $P$
disappear. Furthermore, the factor $1-\alpha^2r^2$ in the function
$\Delta$ is canceled out in the process of finding the limit of the
CM energy (\ref{mcm1}) as $r$ tends to $r_H$.

Let us now to consider the case two massive particles collide near
the acceleration horizon of the black hole. Near the acceleration
horizon $r_A=1/\alpha$, the condition $r\ll
|\frac{1}{\alpha\cos{\theta}}|$ is satisfied only in the region
which is very close to the equatorial plane (i.e., $\theta \sim
 \frac{\pi}{2}$). In the limit $r\rightarrow r_{A}$, we find the CM
energy of the particles becomes
\begin{eqnarray}
E_{
cm_A}^{2}&=&m_{1}^{2}+m_{2}^{2}+\frac{1}{r^2_A+a^2\cos^2\theta}\bigg[(m^2_1r^2_A+K_1)
\nonumber\\&\times&\frac{E_{2}- \Omega_{A}L_{2}}{E_{1}-
\Omega_{A}L_{1}}+(m^2_2r^2_A+K_2)
\frac{E_{1}-\Omega_{A}L_{1}}{E_{2}-\Omega_{A}L_{2}}\nonumber\\&-&\frac{2(L_{1}-a\sin^{2}\theta
E_{1})(L_{2}-a\sin^{2}\theta E_{2})}{(1-2\alpha M\cos \theta
+\alpha^{2}a^{2} \cos
^{2}\theta)\sin^{2}\theta}\nonumber\\&-&\frac{2\sigma_{1\theta}\sqrt{\Theta_{1}}\sigma_{2\theta}\sqrt{\Theta_{2}}}{1-2\alpha
M\cos \theta +\alpha^{2}a^{2} \cos ^{2}\theta}\bigg].
\label{eq:alpha_formula}
\end{eqnarray}
Like in the near event-horizon collision, the CM energy in the near
acceleration-horizon collision $E_{cm_A}$ (\ref{eq:alpha_formula})
depends also on the parameters of the black hole and of the
colliding particles and the polar angle $\theta$ of the orbit of
particles. Since the angular velocity $\Omega_A$ and the
acceleration horizon $r_A$ are a function of $\alpha$, one can find
that the dependence of $E_{cm_A}$ on $\alpha$ is different from that
of $E_{cm_H}$. The derivative of $E_{cm_A}$ with respect to $\alpha$
is
\begin{eqnarray}
\frac{dE^2_{cm_A}}{d\alpha}&=&\frac{\cos\theta(a^2\alpha\cos\theta-M)}{P^2(r^2_A+a^2\cos^2\theta)}
\bigg\{\frac{2(L_{1}-a\sin^{2}\theta
E_{1})}{\sin^{2}\theta}\nonumber\\&\times&(L_{2}-a\sin^{2}\theta
E_{2})+\sigma_{1\theta}
\sqrt{\Theta_{1}}\sigma_{2\theta}\sqrt{\Theta_{2}}
\bigg[2-P\nonumber\\&\times&\bigg(\frac{K_1-m^2_1a^2\cos^2\theta}{\Theta_1}
+\frac{K_2-m^2_2a^2\cos^2\theta}{\Theta_2}\bigg)\bigg]
\bigg\}\nonumber\\&+&\frac{1}{r^2_A+a^2\cos^2\theta}\bigg[\bigg
(\frac{E_{1}-\Omega_{A}L_{1}}{E_{2}-\Omega_{A}L_{2}}-\sigma_{1\theta}\sigma_{2\theta}
\sqrt{\frac{\Theta_1}{\Theta_2}}\bigg)\nonumber\\&\times&\frac{dK_1}{d\alpha}+\bigg(
\frac{E_{2}-\Omega_{A}L_{2}}{E_{1}-\Omega_{A}L_{1}}-\sigma_{1\theta}\sigma_{2\theta}
\sqrt{\frac{\Theta_2}{\Theta_1}}\bigg)\frac{dK_2}{d\alpha}\nonumber\\
&-&\frac{2}{\alpha^3}\bigg(m^2_1\frac{E_{2}-\Omega_{A}L_{2}}{E_{1}-\Omega_{A}L_{1}}
+m^2_2\frac{E_{1}-\Omega_{A}L_{1}}{E_{2}-\Omega_{A}L_{2}}\bigg)
\nonumber\\&+&\frac{2a(E_1L_2-E_2L_1)}{\alpha(1+\alpha^2a^2)}
\bigg(\frac{m^2_1}{(E_{1}-\Omega_{A}L_{1})^2}
\nonumber\\&-&\frac{m^2_2}{(E_{2}-\Omega_{A}L_{2})^2}\bigg)\bigg]
+\frac{2(E^2_{cmA}-m^2_1-m^2_2)}{\alpha(r^2_A+a^2\cos\theta)}.\label{dEcA}
\end{eqnarray}
Obviously, the rate of change $\frac{dE^2_{cm_A}}{d\alpha}$ is
determined by the black hole's parameters ($M$, $a$, $\alpha$), the
particle's parameters ($E_i$, $L_i$, $m_i$) themselves and the
particle orbital parameters. Although in the equatorial plane the
quantity $\cos\theta=0$ and the Carter's constants $K_1$, $K_2$ do
not depend on the acceleration $\alpha$, we find that the CM energy
of the particles in the near acceleration-horizon collision still
depends on the acceleration $\alpha$ of the black hole. This
property of the CM energy $E_{cmA}$ is different from that of
$E_{cmH}$ in the near event-horizon collision.

From Eq.(\ref{eq:alpha_formula}), it seems that we can obtain the
arbitrarily large CM energy $E_{cmA}$ in the near
acceleration-horizon collision if the angular momentum of either
particle $L_i$ is fine-tuned to the critical angular momentum
$L_{icA}=\frac{E_{i}}{\Omega_{A}}=\frac{E_{i}(1+a^2\alpha^2)}{a\alpha^2}$.
However, within a small range near the acceleration horizon, i.e.,
$r=r_A-\delta$, we find that the effective potential for the massive
particle with $L=L_{cA}$  has a form can be expressed as
\begin{eqnarray}
V_{\rm eff}|_{L\rightarrow L_{cA}}&\sim&
\frac{2\alpha(r_A-r_H)(r_A-r_C) }{a^2(1+a^2\alpha^2\cos^2\theta)}
\bigg[(1+a^2\alpha^2\cos^2\theta)\nonumber\\&\times&\frac{E^2}{P\sin^2\theta}
+a^2\alpha^2\bigg(\frac{\Theta}{1+a^2\alpha^2\cos^2\theta}
+m^2\bigg)\bigg]\delta\nonumber\\&+&\mathcal{O}(\delta^2).
\end{eqnarray}
It is always non-negative since $\delta>0$ and $r_A\geq r_H\geq r_C$
in the physical region of the black hole. Combining with
Eq.(\ref{eq:mechanical_energy1}), one can obtain that the particle
with $L=L_{cA}$ can not reach the acceleration horizon since
$(\frac{dr}{d\lambda})^2<0$ in this case. This means that the
arbitrarily high CM energy can not obtained for the particles
colliding near the acceleration horizon.

\section{The high-velocity collision belts in extremal accelerating
and rotating black hole spacetime}

In this subsection, we will focus only on the collision of two
particles with an arbitrarily high CM energy and then probe the
effects the acceleration $\alpha$ of the black hole on the
high-velocity collision belts. As in Ref.\cite{TM1,Liu2}, we here
concentrate our attention only on the direct collision and LSO
collision scenarios in the extremal black hole.

From Eqs. (\ref{eqR}), (\ref{eqThetaRP}) and (\ref{eqTheta}), we
find that $R(r_H)=0$ and $R'(r_H)=0$ are held naturally for the
critical particles in the extremal accelerating and rotating black
hole spacetime. Moreover, the quantity $R''(r_H)$ in this case
becomes
\begin{eqnarray}
R''(r_H)=2\bigg[4M^{2}E^{2}-(m^2M^2+K)(1-M^{2}\alpha^{2})\bigg].
\end{eqnarray}
Since $(1-M^{2}\alpha^{2})>0$, the condition $R''(r_H)\geq0$ leads
to
\begin{eqnarray}
K\leq \frac{4M^{2}E^{2}}{1-M^{2}\alpha^{2}}-m^2M^2.
\end{eqnarray}
Combing with Eq.(\ref{eqKth}), we have
\begin{eqnarray}
&M^{2}&\bigg[\frac{E^{2}(1+\cos^{2}\theta)^2}{(\alpha
M\cos\theta-1)^{2}(1-\cos^{2}\theta)}+m^2\cos^2\theta\bigg]\nonumber\\&\leq
& K\leq\frac{4M^{2}E^{2}}{1-\alpha^{2}M^{2}}-m^2M^2,
\end{eqnarray}
where $a=M$ and $L=L_c=2ME$ have been used. This inequality can be
simplified further as
\begin{eqnarray}
&&m^2\alpha^{2}M^2(1-\alpha^{2}M^2)\cos^6 \theta-2m^2 \alpha
M(1-\alpha^{2}M^2)\cos^5
\theta\nonumber\\&&-[(1+3\alpha^{2}M^2)E^2-m^2(1-\alpha^2M^2)]\cos^{4}\theta\nonumber\\
&&+8\alpha
ME^2\cos^{3}\theta-(6E^2+m^2\alpha^{2}M^2)(1-\alpha^{2}M^2)\cos^{2}\theta\nonumber\\&&
-2\alpha
M[4E^2-m^2(1-\alpha^{2}M^2)]\cos\theta\nonumber\\&&+(3+\alpha^2M^2)E^2-m^2(1-\alpha^2M^2)
\geq0.\label{ine1}
\end{eqnarray}
Obviously, this inequality depends on the acceleration $\alpha$ of
the black hole \cite{TM1}. As $\alpha=0$, we find that it reduces to
results obtained in the usual Kerr black hole spacetime. However,
for $\alpha\neq0$, this inequality is impossible to be solved
analytically. So we must rely on the numerical calculation to obtain
the range of the value of $\cos\theta$ and then probe the effects
the acceleration $\alpha$ of the black hole on the high-velocity
collision belts in this black hole spacetime.

\begin{figure}[htbp]
\includegraphics[width=0.4\textwidth]{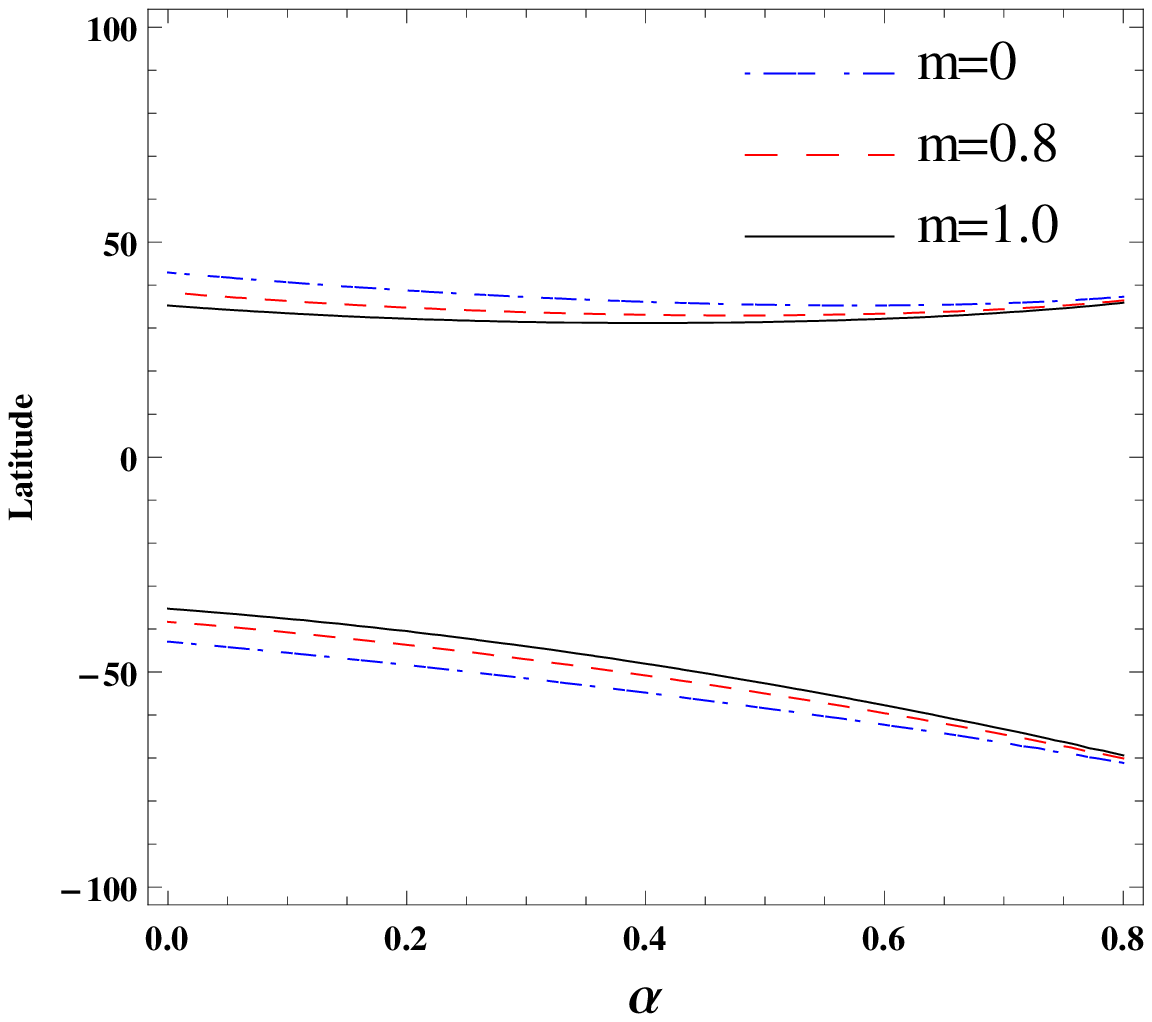}
\includegraphics[width=0.4\textwidth]{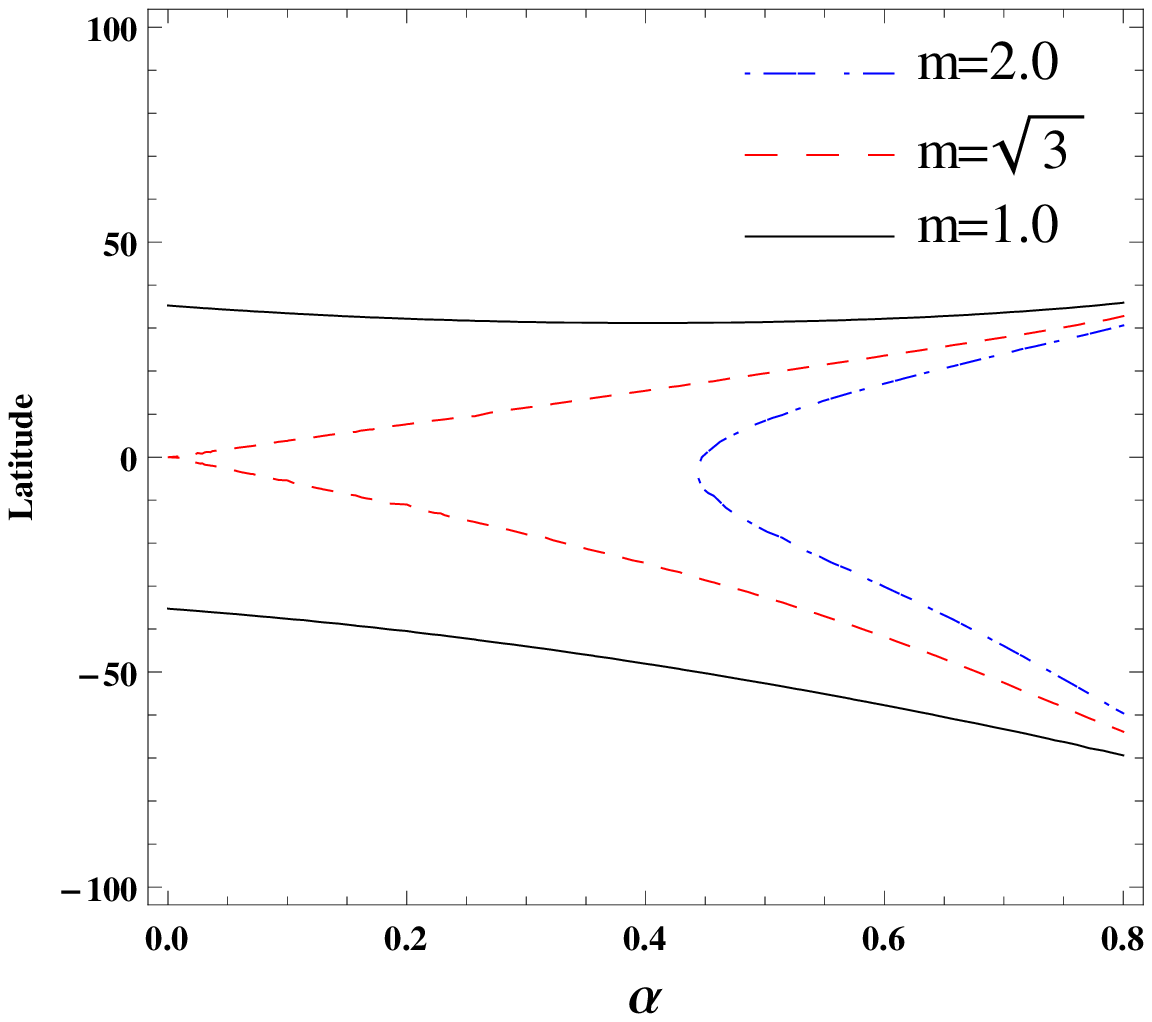}\label{fg:latitude}
\caption{The allowed range of the latitude of high-velocity
collision belts in the extremal accelerating and rotating black
hole. The left and the right are for the particles satisfied
$E^2\geq m^2$ and $E^2\leq m^2$, respectively. We set $M=1$ and
$E=1$.}
\end{figure}

\begin{figure}[htbp]
\includegraphics[width=0.4\textwidth]{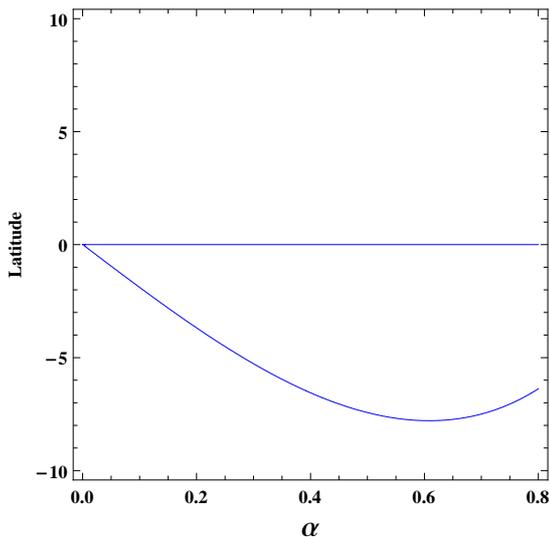}
\caption{The allowed range of the latitude of the high-velocity
collision belts for the critical particles with $(3+\alpha^2M^2)
E^2=(1-\alpha^2M^2) m^2$ in the extremal accelerating and rotating
black hole. respectively. We set $M=1$ and $E=1$.}
\end{figure}

In Fig.(1), we plot the range of the latitude of the orbital plane
of the critical particles in the extremal accelerating and rotating
black hole. It is shown that the high-velocity collision belt is no
longer symmetric with respect to the equatorial plane in this black
hole spacetime. With increase of the acceleration $\alpha$, the
highest absolute value of the south latitude of the orbital plane
increases. But for the particles locating in the northern
hemisphere, the highest allowed latitude value first decreases and
then increases if $3E^2>m^2$,  and increases monotonically if
$3E^2\leq m^2$.  In the extremal Kerr black hole case \cite{TM1},
the high-velocity collision belt of the particles with $3E^2=m^2$
are limited only on the equatorial plane. However, in this
accelerating black hole spacetime,  the presence of the acceleration
$\alpha$ makes the high-velocity collision belt extend to the high
latitude region. Moreover, we can find that the high-velocity
collision belt exists only in the southern hemisphere for the
critical particles with $(3+\alpha^2M^2)E^2=(1-\alpha^2M^2)m^2$ in
this spacetime. The width of the belt first increases and then
decreases with the acceleration $\alpha$, which is shown in Fig.(2).
Furthermore, we find that for the critical particle with the
arbitrary mass $m$ the high-velocity collision belt always exist if
the acceleration $\alpha$ is larger than the critical value
$\alpha_c$, i.e.,
\begin{eqnarray}
\alpha^2M^2\geq\alpha^2_cM^2=\frac{m^2-3E^2}{m^2+E^2}\label{calf}.
\end{eqnarray}
For the particles with $m^2<3E^2$, the condition (\ref{calf}) always
is valid since the right-hand side of the inequality is negative. As
$m^2>3E^2$, there exist a nonzero $\alpha_c$, which increases with
the mass $m$ of the critical particle and reaches its upper limit
$\alpha^2_cM^2=1$ as the mass $m$ tends to infinite. Moreover, we
also find that the high-velocity collision belt becomes more narrow
for the particles with larger mass.

\section{Summary}

In summary, we have studied the collision of two geodesic particles
in the accelerating and rotating black hole spacetime and probe the
effects of the acceleration of the black hole on the CM energy of
the colliding particles and on the high-velocity collision belts in
the limit $r\ll |\frac{1}{\alpha\cos{\theta}}|$. It is shown that
the CM energy depends not only on the black hole parameters and the
colliding particles parameters, but also on the position of the
orbital plane of the colliding particles. Moreover, we also find
that there exists some difference between the properties of the CM
energy in the near event-horizon collision and those of in the near
acceleration-horizon collision. For the particles moving in the
equatorial plane, the CM energy $E_{cm_H}$ in the near event-horizon
collision is independent of the acceleration of the black hole, but
the CM energy $E_{cm_A}$ in the near acceleration-horizon collision
is affected by $\alpha$. The unbound CM energy can be obtained in
the near event-horizon collision if either of two particles has the
critical angular momentum $L_{cH}$, but it is impossible to obtain
in the near acceleration-horizon collision since the effective
potential is positive for the colliding particles with the critical
angular momentum $L_{cA}$.

Finally, we have disclosed the effects of the acceleration $\alpha$
on high-velocity collision belts. Due to the presence of the
acceleration $\alpha$, the high-velocity collision belt is no longer
symmetric with respect to the equatorial plane. With increase of the
acceleration $\alpha$, the highest absolute value of the south
latitude of the orbital plane increases. But for the particles
locating in the northern hemisphere, the highest allowed latitude
value first decreases and then increases if $3E^2>m^2$,  and
increases monotonically if $3E^2\leq m^2$. Moreover, we find that
for the critical particle with the arbitrary mass $m$ the
high-velocity collision belt always exist if the acceleration
$\alpha$ satisfies a critical condition $\alpha>\alpha_c$. For the
particles with $m^2<3E^2$, the critical condition always is valid.
But for the particles with $m^2>3E^2$, there exist a nonzero value
of $\alpha_c$, which increases with the mass $m$ of the critical
particle. Moreover, we find that the high-velocity collision belt
becomes more narrow for the particles with larger mass. Our results
show that the acceleration yields richer effects on the collision of
particles.

\begin{acknowledgement}
This work was  partially supported by the NCET under Grant
No.10-0165, the PCSIRT under Grant No. IRT0964 and the construct
program of key disciplines in Hunan Province. J. Jing's work was
partially supported by the National Natural Science Foundation of
China under Grant No. 10935013; 973 Program Grant No. 2010CB833004.
\end{acknowledgement}

\end{document}